\documentclass[final,a4paper,11pt,3p,twocolumn,times]{elsarticle}

\usepackage{graphicx}
\usepackage{amsmath, amssymb, amsfonts}
\usepackage[colorlinks,citecolor=blue]{hyperref}
\usepackage[american]{babel}
\usepackage{fancyhdr}
\begin{document}

\title{Suryakala-Nusantara: Documenting Indonesian Sundials}

\author[itb]{R. Priyatikanto\corref{cor1}}
\ead{rpriyatikanto@students.itb.ac.id}
\cortext[cor1]{Corresponding author}
\address[itb]{Prodi Astronomi Institut Teknologi Bandung, Jl. Ganesha no. 10, Bandung,\\ Jawa Barat, Indonesia 40132}

\journal{Proceeding of the Indonesian Astronomy Ascossiation (HAI) Seminar 2013}

\hyphenation{}

\begin{abstract}
Sundial is the ancient or classic timekeeper device, especially prior to the invention of mechanical clock. In the classical Islamic civilization, the daily movement of the Sun becomes main indicator of praying time, which can be deduced using sundial. This kind of device probably permeated to Indonesia during the Islamic acculturation. Since then, the development of astronomical knowledge, technology, art and architectural in classical Indonesia are partially reflected into sundial. These historical attractions of sundial demand comprehensive documentation and investigation of Indonesian sundial which are rarely found in the current literatures. The required spatial and temporal information regarding Indonesian sundial can be collected by general public through citizen science scheme. This concept may answer scientific curiosity of a research and also educate the people, expose them with science. In this article, general scheme of citizen science are discussed, its application for sundial study in Indonesia is proposed as Suryakala-Nusantara program.
\end{abstract}

\maketitle

\section{Introduction}
\label{sec1}
A sundial, in its broadest sense, is any device that uses the motion of the apparent sun to cause a shadow or a spot of light to fall on a reference scale indicating the passage of time. Although the oldest sundial was created by Egyptians in 1500 BC \cite{mayall38}, the existence of this device in Islamic civilization came later. Muslim inherited sundial system from the Greeks, who have strong tradition of sundial, during the conquering era around the seventh century. Sundials are then developed and utilized for religious purpose, to indicate the time of midday (\emph{zuhr}) and afternoon (\emph{ashr}) prayer. A number Islamic scientist, such as \emph{Habash al-Hasib} and \emph{Thabit ibn Qurra}, were born and continued the development of sundial knowledge and design in medieval ages \cite{berggren01}.

\begin{figure*}[!t]
\centering
\includegraphics[width=\textwidth]{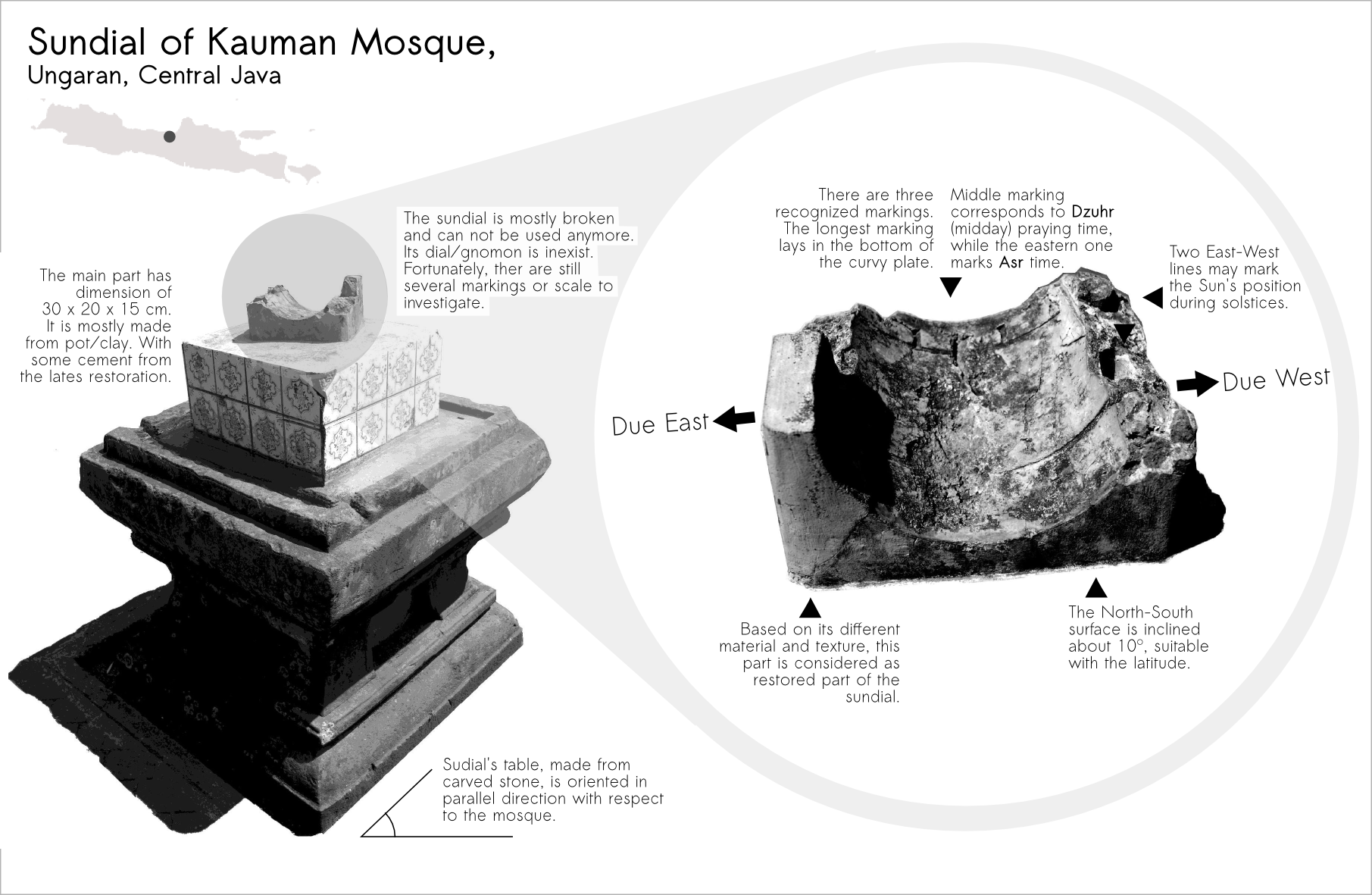}
\caption{Infographic of a sundial that lays in the front rear of Kauman Mosque in Ungaran, Central Java. It is severly damaged with no dial/gnomon to cast a shadow. From the existing features, it can be deduced that this equatorial sundial is a special sundial for praying-timekeeping only. (Photographed by the author)}
\label{kauman}
\end{figure*}

Islamic concepts and its following civilization were then come to Indonesia in fourteenth century mainly through trading activities. Since then, the Islamic ideology and knowledge were widely spread in the Indonesian archipelago. In line with this process, sundial as timekeeper was brought to the archipelago. A large number of mosques are built with different architectural identitiy; some of them are accompanied by sundial.

Unfortunately, the study of the sundial establishment and utilization in Indonesian Islamic civilization has not been precisely conducted and even overshadowed by the studies of the mosque architecture which are more common in literatures. Adequate documentation regarding this timekeeper is hard to find. We do not really know which one is the oldest sundial in Indonesia or even where to find them. Because of its role as timekeeper has been altered by the presence of mechanical and digital clock, sundial gains less attention these days. There are not a few numbers of sundials that have been weathered by time, one of which is a sundial or bancet in Masjid Kauman Ungaran, Jawa Tengah (\autoref{kauman}).

In contrary, similar documentation and research are conducted overseas. Among them are Kim et al. \cite{kim10} and Kim \& Lee \cite{lee11} who documented numbers of Korean sundials which are established in 14th century. The main points of these studies are to understand how the sundials work and to restore some of the defective sundials which they considered as valuable relics. Ferrari \cite{ferrari10}, who registers himself as \emph{North America Sundial Society} (NASS), wrote about medieval sundials constructed by Ottoman in Northern Italy. This kind of research provides an outlook regarding astronomy knowledge, technology, and architectural development of the people at that time. Beside that, the spreading pattern of sundial designs become an important issue to be addressed.

The urgency of sundial documentation and investigation can be responded by utilization of citizen science approach. General public can be actively involved in the documentation of sundials which are distributed on a vast area of Indonesian archipelago and a broad range of time.

This article discusses about the implementation of citizen science scheme in the form of Suryakala-Nusantara. The discussion is limited to the realm of Islamic sundial, though the scope of Suryakala-Nusantara program can be expanded, involving more ancient sundials.
By involving as many as people and exploiting the flexibility of the information sharing through the internet, Suryakala-Nusantara aims to record and document the existence of a sundial in Indonesia. The resulting data/documentation can be used for further studies in many fields such as history, art, architecture, and ethno astronomy.

General explanation about citizen science approach will be given in Section 2, followed by the implementation of the concept into Suryakala-Nusantara program in Section 3, while its on-line database scheme is mentioned in Section 4. Several issues and opportunities will be discussed in the last section (5).

\section{Leaning on Citizen Science}
\label{sec2}
Citizen science can broadly be defined as the involvement of general public or volunteers or non-expert in scientific activities \cite{dickinson10}. This kind of collaboration in astronomy has an equally impressive history. In 1874 the British government funded the Transit of Venus project to measure the Earth's distance to the sun engaging the existing amateur astronomers to support data collection all over the globe. In 1932, British Trust for Ornithology was founded in order to harness the efforts of amateur birdwatchers for the benefit of science and nature conservation \cite{silvertown09}.

However, a new era of citizen science is just rising in the expansion of today's information systems \cite{silvertown09}. Public involvement through citizen science can be categorized into two leading branches, namely data collection and data processing. Numbers of biodiversity researches rely upon citizen science for data collection, for example \mbox{e-Bird}\footnote{\url{www.ebird.com}} which is endorsed by \emph{The Cornell Lab of Ornithology} (CLO) in 2002. On the other hand, \emph{Galaxy Zoo}\footnote{\url{www.galaxyzoo.org, started in 2007}} becomes an example of citizen science project with data processing modus, public are encouraged to access Sloan Digital Sky Survey (SDSS) image of galaxies and do some classifications. In Indonesia, a birdwatcher community called \emph{Indonesian Ornithological Union}\footnote{\url{kukila2004.wordpress.com}} becomes the one example of established citizen science project with several reports/publications produced.

Research project may gain benefits by adopting citizen science scheme since it encompasses a broad scope of space and time. Moreover, citizen science is not only serves positive impact on the scientific research and discovery, but also on the science literacy of the general public \cite{bonney09}. Engagement of the public in citizen-science-based program increases the awareness of science issues and development or the scientific processes that shape the whole human knowledge.

However, this special approach requires a slightly different design when compared to regular research plan. Bonney et al. \cite{bonney09} proposed a basic model for the development of citizen science programs, especially for massive data collection. The development of this model is mainly based on the CLO activities, converges to the following steps:
\begin{itemize}
\item\textbf{Choose scientific questions} for which data collection relies on basic skills of the common participants. More training or supporting materials are required to prepare the participants for higher level involvement.
\item\textbf{Form a team of scientist, educator, engineer, or evaluator team}. The whole chain-process of receiving, archiving, analysing, visualizing, and disseminating project data request a team of multi-discipline people.
\item\textbf{Develop, test, and refine protocols, data forms, and educational support materials}. These factors are needed to ensure the quality of the publicly collected data. Right destination is reached through the right way.
\item\textbf{Recruit participants} through publication via various media such as printed media, e-mail, social media, or workshop at conferences of potential participants or their leaders.
\item\textbf{Train participants} to provide them with sufficient knowledge and skills for data collection or analysis.
\item\textbf{Accept, edit, and display data}. The achieved data need to be available for further analysis, not only for professional scientists, but also the general public.
\item\textbf{Analyse and interpret data}. Raw data from the public need to be analysed to get conclusive points as planned earlier. Here, the professional scientist takes the leading role of the research.
\item\textbf{Disseminate results} to the scientific community and the public. Each segment demands its proper media, e.g. scientific journals or online article for the public.
\item\textbf{Measure outcomes} in order to evaluate the achievement of the project.
\end{itemize}

These steps become inspiring model for the Suryakala-Nusantara establishment.

\section{Composing Suryakala-Nusantara}
\label{sec3}
Suryakala-Nusantara will be formulated according to the model of Bonney et al. \cite{bonney09}. Additional variables (such as funding) are need to taken into consideration.

As previously mentioned, the main objective of Suryakala-Nusantara program is to map the Indonesian sundials, especially those are related to Islamic culture and civilization. The main question to be answered in this research is how the distribution of archipelagic sundials in the dimension of space and time. For this purpose, scientific protocol should includes: (1) take picture clearly, (2) record the location, and (3) find the establishment time of the sundial. Then, the contributor uploads the acquired data to the database of Suryakala-Nusantara.

Sundial picture can be captured using various devices from cell phones to SLR (single-lens reflex) camera which are widely distributed. There are several things to be considered during the photo shoot in order to provide all possible information. Size, orientation, shape, scale and marking detail are 5 informations that should be included in the picture.

The contributor should aware the location of the sundial, at least the name of the mosque where it lies, the area/city and the province. Higher spatial accuracy, e.g. precise geographic coordinate of the sundial, will be required when the case studies are conducted upon particular sundials.

The last protocol is the most challenging step since the date of mosque or sundial establishment is not always present explicitly. Interviewing the elders or scholars becomes an alternative way to obtain the appropriate temporal information.

To support these protocols, public or contributors need to get educational supports which contain basic theory sundial, types, and how the sundial works. Nontechnical aspects, such as art, architecture, and history of general sundial also become subject of the supporting materials.

\section{Compiling Suryakala-Nusantara}
\label{sec4}
This citizen science program will be enriched by the digital data exchange and done via internet. The free domain website, for instance the wordpress domain,   can be employed as the basis homepage in this project. Although it can accommodate every basic information about the Suryakala Nusantara project, in this case the fundamental understanding, scientific protocol, and also education materials. But, the storage capacity of the free domain website is not so large in quantity.

Picture sharing site like flickr.com provide the reliable alternative. One account, both individual or in group, registered in this site are allowed to upload to 1 TB without any charge. The picture description form and discussion forum are also available to simplify the data validation process. Beside, flickr.com is also well-recognized for its scientific discoveries, e.g. newly found species\cite{perkins12}.

Readers can visit \url{www.flickr.com/groups/ suryakala-nusantara/} for contribution or further information.

\section{Perspectives}
The Indonesian classical sundial is a valuable cultural heritage that offers indicator of architectural, technology, art and astronomical knowledge development in the archipelago. It is our task to know, investigate and conserve that heritage. Nevertheless, there are many things to do because of the documentation of Indonesian sundial is rarely found in the literatures.

Citizen science scheme provides a new opportunities in order to transform the physical existence of the sundial into (at least) more secure document. Because of its broad range in space and time domain, it is suitable to be applied in Indonesia through Suryakala-Nusantara program. After being launched in HAI seminar, this program will be promoted widely through the blooming social media.

However, there are several emerging issues related to the data-collection-based citizen science \cite{silvertown09,webb10}:
\begin{itemize}
\item Coordination between the participants need to be conducted in order to ensure the achievement of the main research objectives.
\item Personal and social motivation need to be triggered and maintained from the strong and publicly attractive research backgrounds.
\item Data validation or quality controls become a crucial process to consider for significant scientific product.
\item Retention of the core team and participants. Most of the citizen-science-based researches deal with a long period of time. It is usual that citizen science program brings forth a new community, motivated by discoveries.
\item More funding is needed to accomplish deeper research, especially for in-situ case study of particular sundial.
\end{itemize}
Suryakala-Nusantara that relies heavily on citizen science needs to consider and overcomes these issues.

\section*{Acknoledgements}
The author would like to thank Prof. Dr. Suhardja D. Wiramihadja, Dr. Premana W. Premadi, Dr. Taufiq Hidayat, Dr. Mahasena Putra for the discussions and valuable insights.

\section*{References}
\bibliographystyle{plainnat}
\bibliography{suryakala}
\end{document}